\pdfoutput=1
\documentclass[fleqn,usenatbib]{mnras}
\usepackage{mathtools}
\usepackage{ragged2e}
\usepackage{aas_macros}

\usepackage{graphicx}
\usepackage{amsmath}	
\usepackage{amssymb}	
\usepackage[caption=false]{subfig}
\usepackage[usenames,dvipsnames]{xcolor}
\usepackage[utf8]{inputenc}
\usepackage{color}
\usepackage{hyperref}
\usepackage[normalem]{ulem} 
\usepackage{physics}
\usepackage{multirow}
\usepackage[capitalise]{cleveref}

\usepackage{enumitem}
\usepackage{comment}
\usepackage{bm}
\hypersetup{
  colorlinks   = true, 
  urlcolor     = blue, 
  linkcolor    = blue, 
  citecolor   = blue 
}

\usepackage{soul}

\title[Hybrid disc Geometry of BH-XRBs]{Hybrid disc geometry for shocked accretion flows: Unveiling QPOs in black hole X-ray binaries}

\author[Singh et al]{Monu Singh$^{1}$\thanks{E-mail: monu18@iitg.ac.in},
	Sudip Kumar Garain$^{2}$\thanks{E-mail: sgarain@iiserkol.ac.in}, 
	Santabrata Das$^{1}$\thanks{E-mail: sbdas@iitg.ac.in} \\
	$^{1}$Department of Physics, Indian Institute of Technology Guwahati, Guwahati, 781039, India.\\
	$^{2}$Department of Physical Sciences and Center of Excellence in Space Sciences India, Indian Institute of Science
    Education and \\ Research Kolkata, Mohanpur, West Bengal, India
}

\begin{document}
	\label{firstpage}
	\maketitle
\date{\today}
	
\begin{abstract}
    We investigate the efficacy of semi-analytical global accretion solutions in capturing the flow properties observed in two-dimensional numerical simulations of shocked accretion onto black holes. A comparative analysis reveals that no single disc geometry adequately matches the simulation profiles across the entire radial domain. The pre-shock region exhibits closer agreement with the conical disc geometry, while the post-shock region is better described by the vertical equilibrium disc, where enhanced thermal pressure leads to substantial vertical expansion. Motivated by these complementary behaviours, we introduce a hybrid disc geometry in which the pre-shock flow follows the conical solution and the post-shock flow attains vertical equilibrium. This hybrid model satisfactorily reproduces both dynamical and thermodynamical properties of shocked accretion flows with the predicted Mach number and temperature profiles closely matching the simulations and the inferred shock location differing by $\sim10\%$. Within this framework, we delineate the shock parameter space spanned by the energy ($\mathcal{E}$) and angular momentum ($\lambda$) of the flow for weakly and rapidly rotating black holes and investigate the possible origin of Quasi-periodic Oscillations (QPOs) in black hole X-ray binaries (BH-XRBs). We constrain flow parameters that reproduce observed QPO centroid frequencies ($\nu_{\rm QPO}$) demonstrating that oscillations of the shock front provide a self-consistent mechanism for both low and high frequency QPOs. Extending the analysis to ten Galactic BH-XRBs, we demonstrate that the observed $\nu_{\rm QPO}$ are reproduced within physically plausible parameter ranges, which establishes shocked global accretion solutions as a potentially compelling framework for interpreting accretion driven temporal variability.
\end{abstract}


\section{Introduction}\label{sec:intro}

Accretion onto black holes is considered to be a fundamental astrophysical process that powers some of the most energetic phenomena observed in the Universe, including active galactic nuclei and X-ray binaries \cite[]{Frank-etal2002,Remillard-McClintock2006,Netzer2013,Yuan-Narayan2014}. The standard thin disc model \cite[]{Shakura-Sunyaev1973,Novikov-Thorne1973} successfully explains the thermal emission from accretion discs, however it fails to adequately describe the innermost regions where advection plays a dominant role. Subsequent studies demonstrated that an initially Keplerian flow can become sub-Keplerian close to the black hole \citep{Chakrabarti1990}, motivating the development of advective and geometrically thick accretion models \citep{Narayan-Yi1994, Narayan-Yi1995}.

Accretion flows onto black holes are inherently transonic in nature. Accordingly, rotating accreting matter from the outer edge of the disc becomes supersonic after crossing a critical point \cite[][and references therein]{Chakrabarti1989,Chakrabarti-Das2004} and depending on the angular momentum, flow experiences centrifugal barrier that leads to the triggering the shock transition provided the Rankine–Hugoniot conditions are satisfied \cite{Landau-Lifshitz1959}. The existence of shocks in black hole accretion flows has been firmly established through numerous semi-analytical \cite[]{Fukue1987,Chakrabarti1989,Lu-etal1999,Becker-Kazanas2001,Das-etal2001a,Fukumura-Tsuruta2004,Chakrabarti-Das2004,Das2007,Aktar-etal2015,Aktar-etal2017,Sarkar-etal2018,Dihingia-etal2018a,Mitra-Das2024,Kumar-etal2025} and numerical studies \cite[]{Molteni-etal1996,Lee-etal2011,Das-etal2014,Okuda-Das2015,Sukova-Janiuk2015,Lee-etal2016,Sukova-etal2017,Okuda-etal2019,Okuda-etal2022,Okuda-etal2023,SDebnath-etal2024,Mao-etal2025,Dihingia-etal2025}. The post-shock region is hot and dense, and plays a crucial role in explaining the spectral and timing properties of accreting systems, including the production of hard X-ray emission and quasi-periodic oscillations (QPOs) \cite[]{Chakrabarti-Titarchuk1995,Chakrabarti-Manickam2000,Chakrabarti-etal2008,Chakrabarti-etal2009,Nandi-etal2012,Radhika-Nandi2014,Iyer-etal2015,Sreehari-etal2020,Majumder-etal2022,Majumder-etal2025}.

It is worth noting that the vertical structure of the accretion disc is commonly described by hydrostatic equilibrium, however, alternative disc height prescriptions, such as constant height and conical disc models, have also been widely employed \citep{Chakrabarti1990, Chakrabarti-Molteni1993, Chakrabarti-Das2001, Giri-etal2010, Bilic-etal2014, Zhou-etal2025}. Although these models are globally equivalent under suitable transformations of the polytropic index \citep{Chakrabarti-Das2001}, their local flow properties and shock characteristics differ significantly. In particular, numerical simulations indicate that no single disc-height prescription can accurately describe the entire accretion flow in the presence of shocks, as the pre-shock region is better represented by a conical geometry, whereas the post-shock region exhibits vertical expansion consistent with the disc structure obtained using the hydrostatic equilibrium prescription \citep{Giri-etal2010}.

Motivated by this discrepancy, we propose a physically motivated hybrid disc geometry in which the pre-shock flow follows a conical disc prescription, while the post-shock flow attains a vertical equilibrium structure due to enhanced thermal pressure. To achieve this, we consider a steady, inviscid, advective flow around a rotating black hole and adopt an effective potential that mimics the spacetime geometry of a Kerr black hole \citep{Dihingia-etal2018}. Within this framework, we compute global transonic solutions featuring standing shocks and demonstrate that the hybrid disc model yields significantly better agreement with numerical simulations compared to single geometry prescriptions. Furthermore, we investigate shock oscillations as a plausible mechanism for the origin of Quasi-periodic Oscillations (QPOs). Employing our model formalism, we reproduce the observed QPO frequencies ($\nu_{\rm QPO}$) by constraining the underlying flow parameters, namely energy ($\mathcal{E}$) and angular momentum ($\lambda$) of the flow, along with shock properties such as the shock radius ($r_s$) and compression ratio ($R$) for several Galactic black hole X-ray binaries (BH-XRBs). 

The paper is organised as follows. Section~\ref{sec:02} outlines the model assumptions and governing equations describing the accretion flow around a rotating black hole. In Section~\ref{sec:03}, we present the critical point analysis and construct shock-induced global accretion solutions. Section~\ref{sec:04} provides a detailed discussion of our results, while Section~\ref{sec:05} explores the astrophysical implications of our formalism in explaining QPOs from BH-XRBs. Finally, we present the conclusions in Section~\ref{sec:06}.

\section{GOVERNING EQUATIONS AND ASSUMPTIONS}\label{sec:02}

We examine the dynamics of low angular momentum, advective accreting flow around a rotating black hole. To capture different geometric configurations of the accretion flow, we consider three axisymmetric and inviscid disc models: (a) Model V, where the flow is in hydrostatic equilibrium along the transverse direction; (b) Model C, characterized by a conical cross-section in the meridional plane; and (c) Model H, where the disc maintains a constant vertical height all throughout. The analysis is carried out in cylindrical coordinates ($r$, $\phi$, $z$). Under these assumptions, the governing fluid equations are formulated using a unit system $M_{\rm{BH}} = G = c = 1$, where $M_{\rm{BH}}$ is the black hole mass, $G$ is the gravitational constant, and $c$ is the speed of light. In this unit system, the length, angular momentum and energy of the flow are measured in units of $r_g = G M_{\rm{BH}}/c^2$, $cr_g$ and $c^2$, respectively. With these considerations, we formulate the set of fluid equations that govern the accretion flow confined at the disc equatorial plane around the rotating black holes and are given by \cite[]{Dihingia-etal2018},

\noindent (i) Radial momentum equation:
\begin{equation}
\label{eqn:radmom}
\upsilon \frac{d\upsilon}{dr}+\frac{1}{h \rho}\frac{dP}{dr}+\frac{d \Phi_{\textrm{e}}^{\textrm{eff}}}{dr} = 0,
\end{equation}
where, $\upsilon$ is the radial velocity, $P$ is the pressure, $\rho$ is the mass density, $h ~[=(\epsilon +P)/\rho]$ is the enthalpy, and $\epsilon$ denotes the internal energy of the flow, respectively. In equation (\ref{eqn:radmom}), $\Phi_{\textrm{e}}^{\textrm{eff}}$ refers the effective gravitational potential that satisfactorily describes the space-time geometry around the rotating black holes at the disc equatorial plane and is given by \cite{Dihingia-etal2018},
$$
\Phi_{\textrm{e}} ^{\textrm{eff}} = \frac{1}{2}\ln\left[\frac{r \Delta}{a_{\textrm{k}}^2 (r+2) - 4 a_{\textrm{k}} \lambda + r^3 -\lambda^2(r-2)}\right],
$$
where $\lambda$ denotes the angular momentum of the flow, $a_{\rm k}$ represents the spin of the black hole, and $\Delta = r^2 - 2 r + a_{\rm k}^2$, respectively.

\noindent (ii) Mass conservation equation:
\begin{equation}
\label{eqn:accrate}
{\dot M} = 2\pi \upsilon \Sigma \sqrt{\Delta} = 4 \pi \upsilon \rho \sqrt{\Delta} r^\beta \left( \frac{Pr}{\rho \mathcal{F}}\right)^{\delta/2}
\end{equation}
Equation (\ref{eqn:accrate}) is obtained from the mass flux conservation equation and it depends on specific disc geometry under consideration. Here, $\dot M$ denotes the mass accretion rate which remains constant throughout the flow, and $\beta$ and $\delta$ are constants. For model V \cite[]{Riffert-Herold1995,Peitz-Appl1997}, $\beta=1$ and $\delta=1$. For Model C \cite[]{Chakrabarti1990}, $\beta=1$ and $\delta=0$. For Model H \cite[]{Chakrabarti1992,Chakrabarti-Das2001}, $\beta=0$ and $\delta=0$. In addition, $\mathcal{F} = \left[(r^2 + a_{\rm k}^2)^2 + 2 \Delta a_{\rm k}^2\right] / \left[\left(1-\lambda \Omega\right) \left( (r^2 + a_{\rm k}^2)^2 - 2 \Delta a_{\rm k}^2 \right) \right]$
and $\Omega ~(= \left[2 a_{\textrm{k}} + \lambda (r - 2)\right] / \left[a_{\textrm{k}}^{2}(r + 2) - 2 a_{\textrm{k}}\lambda + r^3\right])$ is the angular velocity of the flow.

\noindent (iii) Entropy generation equation:
\begin{equation}
\label{eqn:entpy}
\left( \frac{\epsilon+P}{\rho}\right)\frac{d \rho}{d r} - \frac{d \epsilon}{d r}=0.
\end{equation}
To close the set of governing equations (\ref{eqn:radmom}), (\ref{eqn:accrate}), and (\ref{eqn:entpy}), we employ, owing to simplicity, an equation of state (EoS) that relates $\epsilon$, $P$ and $\rho$ of the flow as \cite[]{Chattopadhyay-Ryu2009},
\begin{equation}
	\label{eqn:eos}
	\epsilon = \rho + \frac{P}{\Gamma - 1},
\end{equation}
where $\Gamma$ denotes the adiabatic index that remains fixed all throughout the flow. Subsequently, we define the sound speed of the flow as $C_{\rm s}=\sqrt{\Gamma P/(\epsilon+P)}$.

Upon simplifying equations (\ref{eqn:radmom})-(\ref{eqn:eos}), we calculate the gradient of radial flow velocity as,
\begin{equation}
    \label{eqn:dvdr}
    \frac{d\upsilon}{d r} = \frac{\mathcal{N}}{\mathcal{D}},
\end{equation}
where the explicit expression of numerator $\mathcal{N}$ and denominator $\mathcal{D}$ for various disc geometries are given by,
\begin{equation}
\label{eqn:num}
    \begin{aligned}
        \mathcal{N}	 &= \frac{2C_{s}^2}{2 +\delta(\Gamma-1)}\left[\frac{1}{2 \Delta}\frac{d \Delta}{dr} + \frac{\delta}{2} \left(\frac{1}{r} - \frac{1}{\mathcal{F}} \frac{d \mathcal{F}}{d r}\right) + \frac{\beta}{r}\right]\\
        & \quad\quad -\left(\frac{d\Phi_{\rm{e}}^{\rm{eff}}}{d r}\right),
    \end{aligned}
\end{equation}
and,
\begin{equation}
\label{eqn:deno}
\mathcal{D}	= \upsilon - \frac{2 C_{s}^2}{\left[2 +\delta(\Gamma-1)\right]  \upsilon}.
\end{equation}

Next, we express flow temperature in dimensionless form as $\Theta = k_{\textrm{B}} T/m_{\textrm{e}} c^2$, where $T$ is the flow temperature in Kelvin, $k_B$ is the Boltzmann constant, and $m_e$ is the electron mass, respectively. With this, we calculate the gradient of flow temperature as,
\begin{equation}
\label{eqn:dtheta}
\frac{d\Theta}{d r}
= - \frac{2\Theta}{ 2N+\delta}\left[ \frac{1}{ \upsilon} \frac{d \upsilon}{d r}+
\frac{1}{2 \Delta}\frac{d \Delta}{d r} + \frac{\delta}{2}\left(\frac{1}{r} - \frac{1}{ \mathcal{F}} \frac{d \mathcal{F}}{d r} \right)
+  \frac{\beta}{r}\right],
\end{equation}
where $N~\left[=1/(\Gamma-1) \right]$ is the polytropic index of the flow.

\section{Critical Point Analysis and shocked accretion solutions}\label{sec:03}

The accretion flow around a black hole is inherently transonic in nature. At large distances from the black hole, the flow is subsonic, but as it moves inward and approaches the event horizon, it makes sonic transition to supersonic regime. Consequently, the flow must pass through critical point ($r_{\rm c}$) before plunging into the black hole. At the critical point, both numerator $\mathcal{N}$ and denominator $\mathcal{D}$ of equation (\ref{eqn:dvdr}) simultaneously vanish, resulting in an indeterminate form $d\upsilon/d r=0/0$. Imposing the condition $\mathcal{D}=0$ (see equation \ref{eqn:deno}) leads to the Mach number at $r_{\rm c}$ as,
\begin{equation}
\label{eqn:mach}
M(r_{\rm c}) = \frac{\upsilon}{C_{\rm s}} = \sqrt{\frac{2}{2 + \delta(\Gamma - 1)}},
\end{equation}
Enforcing $\mathcal{N}=0$ (see equation \ref{eqn:num}) then provides an expression for the sound speed at $r_{\rm c}$ as,
\begin{equation}
\label{eqn:11}
\begin{aligned}
C_{\rm s}^{2} (r_{\rm c}) &= \left[ 1+\frac{\delta(\Gamma-1)}{2}\right] \left(\frac{d \Phi_{\rm e}^{\rm eff}}{d r} \right)\\
&\quad \times \left[ \frac{1}{2 \Delta} \frac{d \Delta}{d r}
+ \delta \left(\frac{1}{2 r}
- \frac{1}{2 \mathcal{F}} \frac{d \mathcal{F}}{d r}\right)
+ \frac{\beta}{r} \right]^{-1}.
\end{aligned}
\end{equation}

Since equation (\ref{eqn:dvdr}) becomes indeterminate at $r_{\rm c}$, we apply l$'$H\^{o}pital's rule to compute the radial velocity gradient of the flow. This yields two possible values of the velocity gradient $(d\upsilon/dr)_{\rm c}$ at $r_{\rm c}$. The nature of the critical point is determined by these values: if both are real and of opposite sign, the critical point is classified as a saddle-type; if both are real and share the same sign, it is a nodal-type; and if the values are complex, the point is deemed a spiral-type \cite{Chakrabarti1990, Chakrabarti-Das2004, Das2007}. Note that saddle-type critical points (hereafter critical points) are stable and particularly relevant as the accretion flow can smoothly pass through them while transitioning into the black hole \cite[]{Liang-Thompson1980,Abramowicz-Zurek1981,Kato-etal1993}. Depending on the choice of input model parameters ($\mathcal{E}, \lambda, a_{\rm k}$) and disc geometry, the accretion flow may possess multiple critical points. We refer to the critical point forming close to the event horizon as the inner critical point ($r_{\rm in}$), and the one located farther out as the outer critical point ($r_{\rm out}$). To obtain the global transonic accretion solutions for a set of model parameters, we numerically integrate the governing equations (\ref{eqn:dvdr}) and (\ref{eqn:dtheta}) starting from a critical point ($r_{\rm c}$), and extend the integration both inward toward the event horizon ($r_{\rm H}$) and outward to the outer edge of the accretion disc ($r_{\rm edge}$). By smoothly connecting these two branches, we obtain a global transonic accretion solution \cite{Chakrabarti-Das2004, Das2007,Singh-Das2024}.

The existence of multiple critical points in such solutions not only determines the transonic nature of the flow but also plays a crucial role in enabling discontinuous transitions within the accretion dynamics. In particular, the formation of a shock is a natural consequence of the complex interplay between gravity and centrifugal repulsion. For a shock transition to occur, the accreting matter must possess multiple critical points allowing the flow to undergo a discontinuous transition from supersonic to subsonic branch \cite[]{Fukue1987,Chakrabarti1989,Chakrabarti-Das2004,Das2007}. This scenario arises when the centrifugal force, acting as an effective barrier against radial infall, becomes strong enough to balance the inward gravitational pull. As the supersonic inflow encounters this centrifugal barrier, a shock transition is triggered provided the Rankine–Hugoniot conditions (RHCs) are satisfied \cite[]{Landau-Lifshitz1959}. These conditions require the conservation of mass flux, momentum flux, and energy flux across the shock front. More specifically, the RHCs are: (a) mass flux conservation ${\dot M}_{+}={\dot M}_{-}$, (b) energy flux conservation ${\mathcal{E}}_{+}={\mathcal{E}}_{-}$, (c) momentum flux conservation $\mathcal{H}_{+}[P_{+} + \rho_{+}v^2_{+}] = \mathcal{H}_{-}[P_{-}+\rho_{-}v^2_{-}]$, where `$-$' and `$+$' refer quantities immediately before and after the shock transition and $\mathcal{H}$ represents the local half-thickness of the disc. To estimate the energy flux, we use local specific energy as $\mathcal E(r)=\frac{1}{2} v^2+\log h + \Phi^{\rm eff}_{\rm e}$ which combines kinetic, thermal, and gravitational contributions of the accretion flow.

\begin{figure}
    \begin{center}
        \includegraphics[width=\columnwidth]{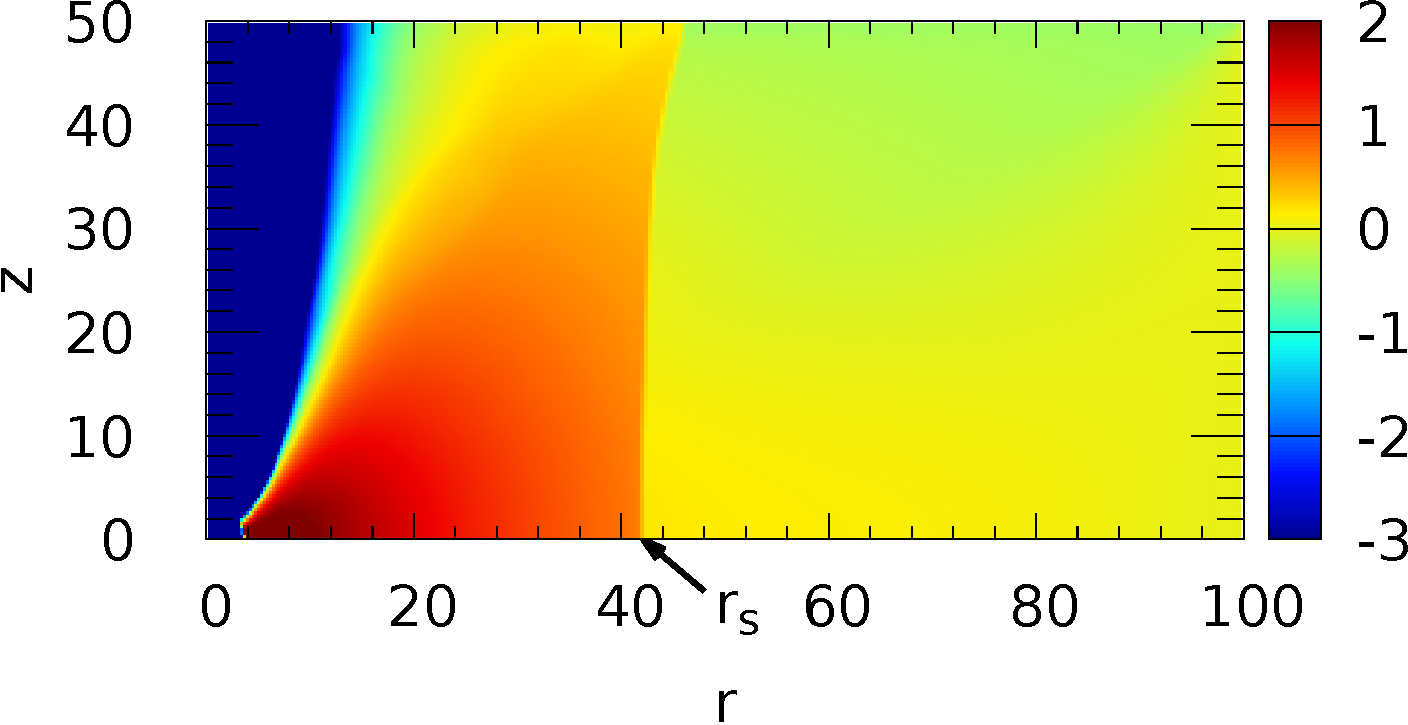}
    \end{center}    
    \caption{Distribution of logarithmic normalized density $\log_{10}(\rho/\rho_{100})$ in the $r$$–$$z$ plane. The pronounced density discontinuity near $r_s \sim 42 r_g$ on the equatorial plane ($z=0$) extending vertically upward indicates the shock surface. Inner critical at $r_{\rm in} \sim 5.08r_g$ is also marked. See the text for details.
    }
    \label{fig:01} 
\end{figure}

Thereafter, we compare our semi-analytical results with numerical simulation. For this purpose, we employ the relativistic hydrodynamics code developed by \citet{Garain-Kim2023}. The simulations solve the ideal hydrodynamical equations in a two-dimensional cylindrical coordinate system $(r,z)$ over the domain $0 \leq r \leq 100$ and $0 \leq z \leq 50$ using a grid resolution of $256 \times 128$ with a common stretching ratio of $1.003$. The computational domain covers only the first quadrant, with reflection symmetry imposed across the equatorial plane. Inflow boundary conditions are applied along $r = 100$ allowing matter to enter the domain through a vertical disc height $\mathcal{H} \sim 50$. It is important to note that the injection height $\mathcal{H}$ employed in the simulations should not be identified with the disc height prescriptions used in the analytical models as the former serves only as a boundary condition, whereas the latter defines the assumed global geometrical structure of the accretion flow. Outflow conditions are set along $z = 50$, while a reflecting boundary is enforced at $r = 0$. To ensure numerical accuracy, the simulations employ a globally second-order scheme. Spatial reconstruction is performed using the van Leer slope limiter \cite[]{Mignone2014}, and temporal evolution is carried out using a two-stage strong-stability-preserving Runge–Kutta method \cite[SSP\_RK2;][]{Shu-Osher1988}. The interfacial fluxes are computed with the Harten–Lax–van Leer (HLL) Riemann solver. The simulation domain is initially filled with a tenuous background medium with a density floor of $\rho_{\rm floor} = 10^{-6}$ and a pressure floor $P_{\rm floor} = 10^{-9}$ to avoid numerical singularities. As the simulation progresses, the inflowing matter rapidly dominates the domain forming a stable shock structure due to interaction with the centrifugal barrier near the black hole. Fig. \ref{fig:01} illustrates the distribution of logarithmic normalized density ($\log_{10} [\rho/\rho_{100}]$) in the $r$$-$$z$ plane at time $t = 50000~r_g/c$, when the simulation reaches a steady state. For this simulation, we adopt the parameters ${\cal E} = 1.0043$, $\lambda = 3.38$, $a_{\rm k} = 0.0$, and $\Gamma = 4/3$. Here, $\rho_{100}$ denotes the density at $r=100~r_g$. The shock surface is evident from the abrupt density jump near $r \sim 42~r_g$ along the disc equatorial disc ($z=0$), which extends vertically upward. The region near the axis remains void of matter due to the non-zero angular momentum of inflow. This feature provides a distinct signature of the flow dynamics and emphasizes the role of angular momentum in governing the accretion geometry.

\section{Results}\label{sec:04}

In this section, we present the results of our investigation into the structure of transonic global accretion flows around black holes for different disc geometries. Using a combination of semi-analytical techniques and high resolution numerical simulations, we examine the dynamical properties of the flow. This allows us to assess the validity of theoretical models under different geometric configurations by directly comparing them with simulation outcomes.

\subsection{Global Transonic Accretion Solutions with Shock}

\begin{figure}
    \begin{center}
        \includegraphics[width=\columnwidth]{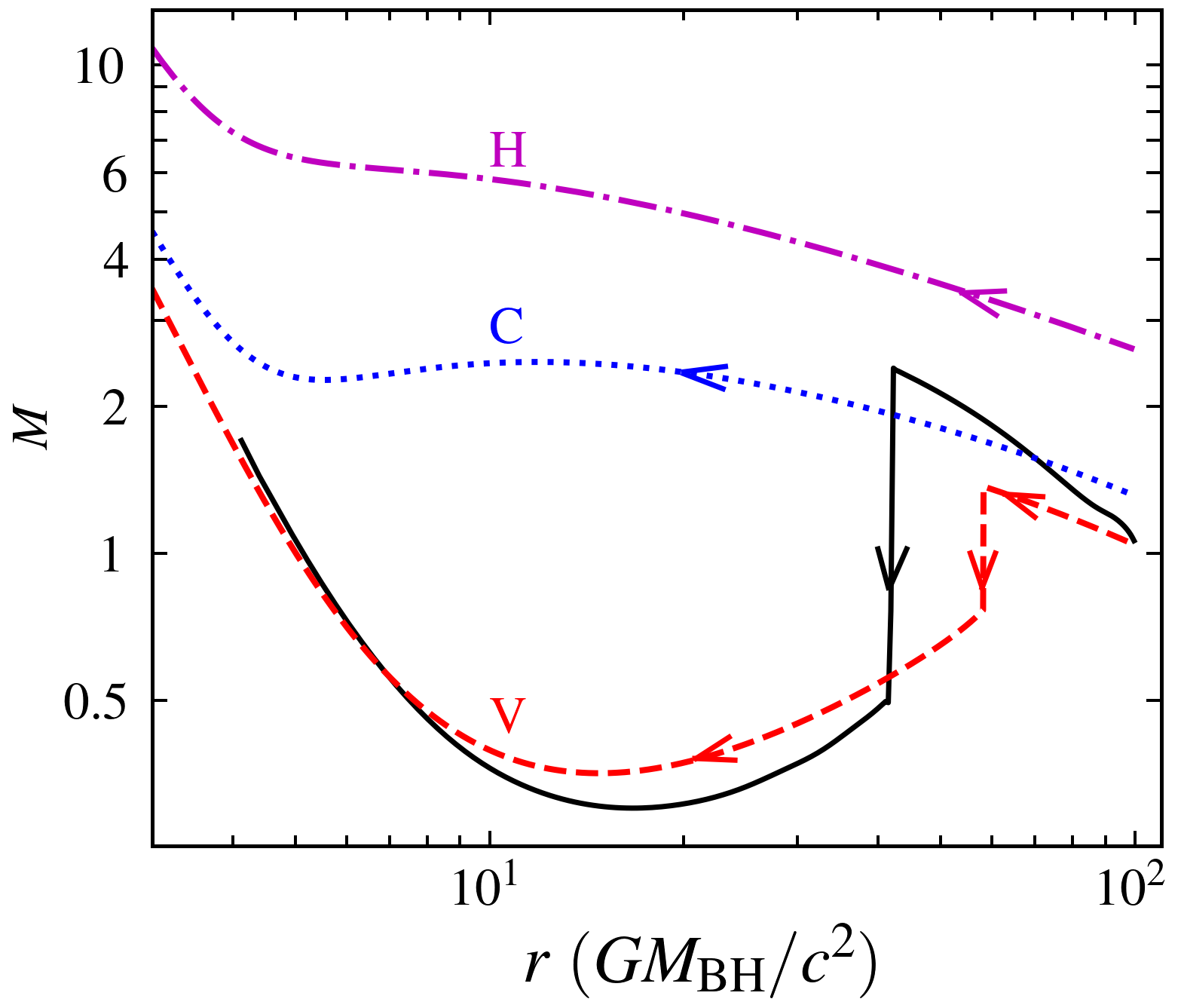}
    \end{center}    
    \caption{Radial variation of the Mach number ($M = \upsilon / C_{\rm s}$) for different disc geometries. The dashed (red), dotted (blue), and dot–dashed (purple) curves correspond to Model-V, Model-C, and Model-H, respectively, while the solid (black) curve represents the simulation result at the disc equatorial plane ($z=0$). The parameters are chosen as $\mathcal{E} = 1.0043$, $\lambda = 3.38$, $\Gamma = 4/3$, and $a_{\rm k} = 0.0$. Vertical arrows denote the locations of shock transitions. See the text for details.
    }
    \label{fig:02} 
\end{figure}

\begin{figure}
    \begin{center}
        \includegraphics[width=\columnwidth]{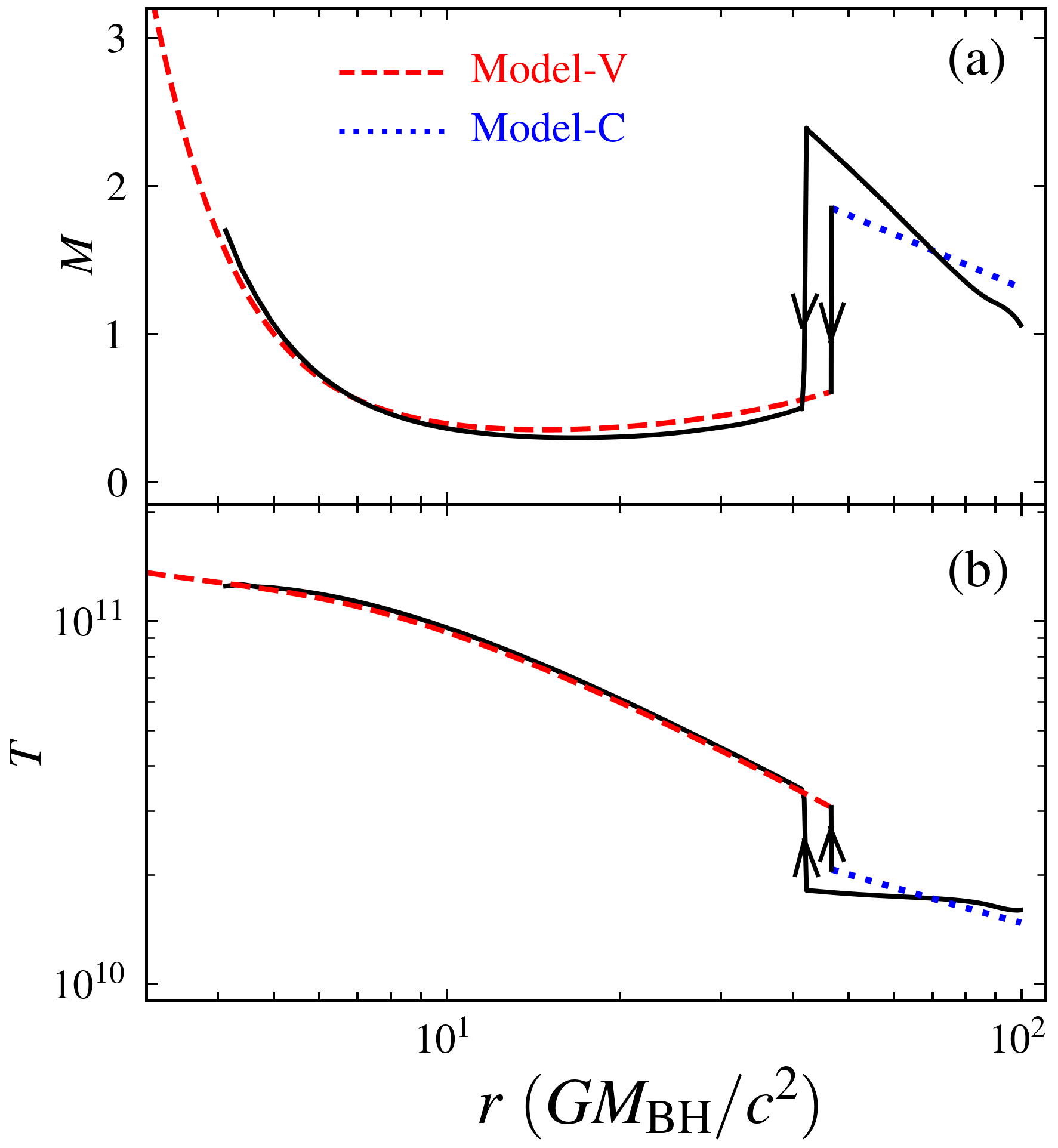}
    \end{center}    
    \caption{(a) Radial variation of the Mach number ($M = \upsilon / C_{\rm s}$) for the hybrid disc geometry. The dotted (blue) and dashed (red) curves denote the components of the hybrid solution obtained from Model-C and Model-V, respectively. The solid (black) curve represents the simulation result as in Fig. \ref{fig:02}. Here, we choose $\mathcal{E} = 1.0043$, $\lambda = 3.38$, $\Gamma = 4/3$, and $a_{\rm k} = 0.0$. Vertical arrows mark the locations of shock transitions. (b) Comparison of the radial variation of temperature ($T$) between the hybrid disc geometry and the corresponding simulation result shown in panel (a). See the text for details.
    }
    \label{fig:03} 
\end{figure}

In Fig.~\ref{fig:02}, we compare the transonic accretion solutions obtained from three different disc geometries. These solutions are computed using an identical set of flow parameters, which are ${\cal E} = 1.0043$, $\lambda = 3.38$, $a_{\rm k} = 0.0$, and an adiabatic index $\Gamma = 4/3$. The figure shows the variation of the Mach number ($M = \upsilon/C_s$) as a function of the radial coordinate $r$. The solutions corresponding to Model-V, Model-C, and Model-H are displayed by the dashed (red), dotted (blue), and dot–dashed (purple) curves, respectively, while the numerical simulation result at the disc equatorial plane ($z=0$) is over plotted using the solid (black) curve. The comparison clearly demonstrates that the three semi-analytical solutions differ substantially across the respective disc geometries, despite being calculated using identical input parameters. This indicates the strong dependence of the transonic flow structure on the assumed disc geometry. More precisely, we find that Model-V exhibits a discontinuous shock transition at $r_{\rm s}=58.24$, whereas both Model-C and Model-H remain entirely smooth and shock-free throughout the entire computed domain including the supersonic region beyond $100\,r_g$. The numerical simulation also shows the presence of a shock at $ r_{\rm s}=42.08$, although its location does not coincide with the shock predicted by Model-V. In the figure, vertical arrows denote the positions of shock transitions, and remaining arrows indicate the inward direction of matter flow as it plunges toward the black hole.

Indeed, the above discrepancy highlights the intrinsic dynamical nature of the simulated flow. It is worth noting that in the simulation, the disc geometry is not prescribed a priori. Instead, the initially unbounded inflowing matter evolves self-consistently into an accretion configuration as the system progresses dynamically. Consequently, the disc structure emerges naturally from the coupled hydrodynamic and gravitational interactions, rather than being constrained by an assumed geometrical prescription. Overall, these differences emphasize the necessity of carefully choosing disc geometry while developing the physically realistic accretion flow models around black holes.

By examining the simulation results, we observe that none of the semi-analytical solutions perfectly replicate the simulation throughout the entire domain. Instead, we observe that the pre-shock branch of the flow broadly agrees with the semi-analytical solution obtained from Model-C, whereas the post-shock branch closely resembles the behaviour predicted by Model-V, where the flow undergoes vertical expansion driven by elevated thermal pressure. Motivated by this, we consider a model of hybrid disc geometry in which the pre-shock flow is described by Model-C and the post-shock flow is prescribed by Model-V. Using this combined configuration, we compute the shock-induced global accretion solution, and the obtained results are presented in Fig. \ref{fig:03} along with the simulation profiles. Fig. \ref{fig:03}(a) shows the radial variation of the Mach number, while Fig. \ref{fig:03}(b) depicts the temperature ($T$) profile with $r$. In both panels, the hybrid model demonstrates a close agreement with the simulation in terms of the Mach number ($M$) and temperature ($T$) profiles. The hybrid solution predicts a shock location at $r_{\rm s}=46.68$, differing by only $\sim 10\%$ from the shock radius obtained in the simulation. This residual discrepancy primarily arises from the different treatments of the vertical flow structure in the semi-analytical and numerical approaches. In the former, the disc geometry is prescribed a priori, whereas in the latter it develops self-consistently from the underlying flow dynamics. We stress that the difference in the shock location between Figs.~\ref{fig:02} and \ref{fig:03} is solely a consequence of adopting the hybrid disc geometry. The strength of the hybrid model lies in its ability to reconcile the distinct geometric characteristics of the pre-shock and post-shock regions while preserving the conservation laws governing the shock transition. Consequently, it provides a more realistic representation of the simulated flow structure and substantially improves the agreement between the semi-analytical solutions and the numerical simulation. We also compute the compression ratio ($R$), which quantifies the density compression across the shock front \cite[][and references therein]{Das2007} and obtain $R=2.48$. Furthermore, as shown in Fig. \ref{fig:03}b, the temperature profile ($T$) predicted by the hybrid model exhibits good agreement with the simulation. This finding evidently supports the physical plausibility and robustness of the proposed hybrid framework.

\begin{figure}
    \begin{center}
        \includegraphics[width=\columnwidth]{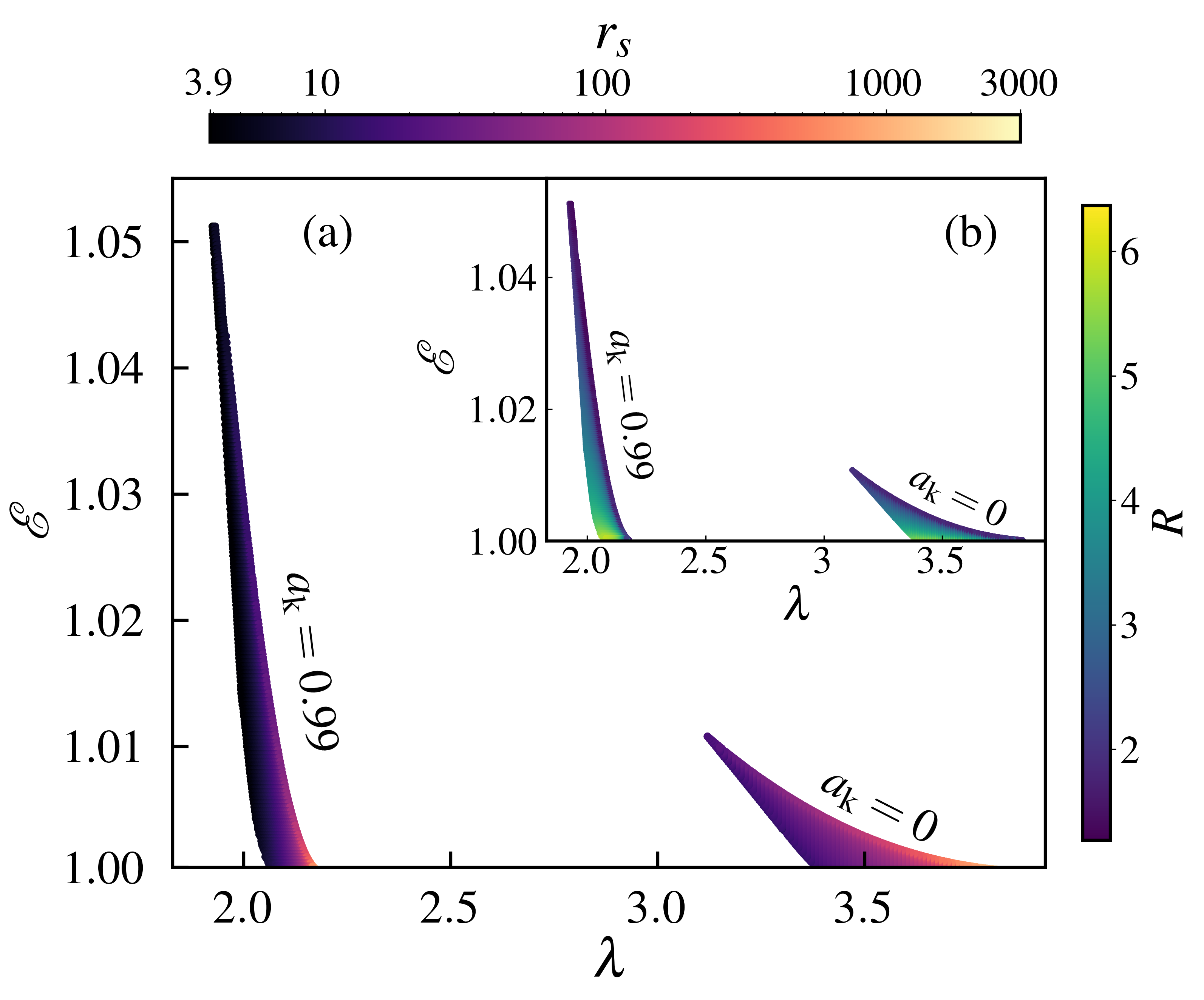}
    \end{center}    
    \caption{Two-dimensional projection of the three-dimensional parameter space spanned by the angular momentum ($\lambda$), energy ($\mathcal{E}$), and shock location ($r_{s}$). Results are shown for weakly rotating ($a_{\rm k} = 0$) and rapidly rotating ($a_{\rm k} = 0.99$) cases. The inset shows the identical $\lambda$–$\mathcal{E}$ parameter space color coded by the compression ratio ($R$). The horizontal and vertical color bars indicate the ranges of $r_{s}$ and $R$, respectively. Here, we choose $\Gamma = 4/3$. See the text for details.
    }
    \label{fig:04} 
\end{figure}

\subsection{Parameter space with hybrid disc height for standing shock}

In this section, we examine the ranges of specific energy ($\mathcal{E}$) and angular momentum ($\lambda$) that admit shock-induced global accretion solutions within the hybrid model framework. To this end, we separate the effective domain of the parameter space on the $\lambda–\mathcal{E}$ plane for standing shocks formed in accretion flows around both weakly rotating ($a_{\rm k}=0$) and rapidly rotating ($a_{\rm k}=0.99$) black holes. The resulting parameter spaces are shown in Fig. \ref{fig:04}, which demonstrate that shocked accretion solutions are not isolated but instead exist over a broad range of $\lambda$ and $\mathcal{E}$. It is worth mentioning that shock transitions occur for flows with comparatively lower angular momentum around rapidly rotating black holes than around weakly rotating ones. This behaviour arises due to spin–orbit coupling embedded into the spacetime geometry around the rotating black holes \cite[]{Sen-etal2022, Das-Chakrabarti2008, Dihingia-etal2019}. In Fig. \ref{fig:04}a, the color scale represents the shock location ($r_s$) with the corresponding color bar indicating the range of standing shock radii. It is evident that, for a given $\mathcal{E}$, shocks generally form at larger radii for flows with relatively higher angular momentum. Moreover, we note that identical shock radius may be obtained for different sets of ($\lambda$, $\mathcal{E}$). We further observe that the black hole spin plays a crucial role in determining the shock properties, with the shock location moving progressively outward as the spin parameter increases for a given set of flow parameters (see Appendix-\ref{app:spin}). In Fig. \ref{fig:04}b, we replot the $\lambda-\mathcal{E}$ plane for standing shock with the compression ratio ($R$) indicated by the color scale, and the adjacent color bar denotes its range. We further notice that nature of shocks becomes stronger ($R \gtrsim 4$) when shock forms closer to the black hole horizon. Such shock-induced accretion solutions have been successful in interpreting the observed spectral and temporal properties of black hole sources \cite[]{Chakrabarti-Titarchuk1995, Chakrabarti-Manickam2000, Mandal-Chakrabarti2005, Nandi-etal2012, Iyer-etal2015, Das-etal2021, Majumder-etal2022, Nandi-etal2024}. In the following section, we demonstrate how these solutions within the hybrid model can be employed to account for the observed QPO frequencies in various black hole systems.

\subsection{Estimation of QPO frequency}

In this section, we examine the properties of standing shocks to establish a connection between shock dynamics and the observed quasi-periodic oscillations (QPOs). It is noteworthy that when the conditions for a standing shock are unfavorable and the entropy of the inner critical point ($r_{\rm in}$) solution exceeds that of the outer critical point ($r_{\rm out}$) solution, the shock front fails to attain a steady configuration and instead exhibits non-stationary behavior \cite[]{Dihingia-etal2019}. Such oscillations arise either from resonance effects, where the post-shock cooling timescale becomes comparable to the infall timescale of the flow \cite[]{Molteni-etal1996}, or from unstable perturbations triggered when viscosity exceeds its threshold limits \cite[]{Lee-etal2011, Lee-etal2016, Das-etal2014, Debnath-etal2024, Debnath-etal2025}. The resulting oscillations of the shock front are generally quasi-periodic in nature and are regarded as the plausible physical origin of the QPOs observed in black hole X-ray binaries (BH-XRBs). 
Therefore, estimating the shock location is of fundamental importance, as it provides a direct link between the dynamics of the accretion flow and the observed timing properties of black hole sources.

\begin{figure}
	\begin{center}
	   \includegraphics[width=\columnwidth]{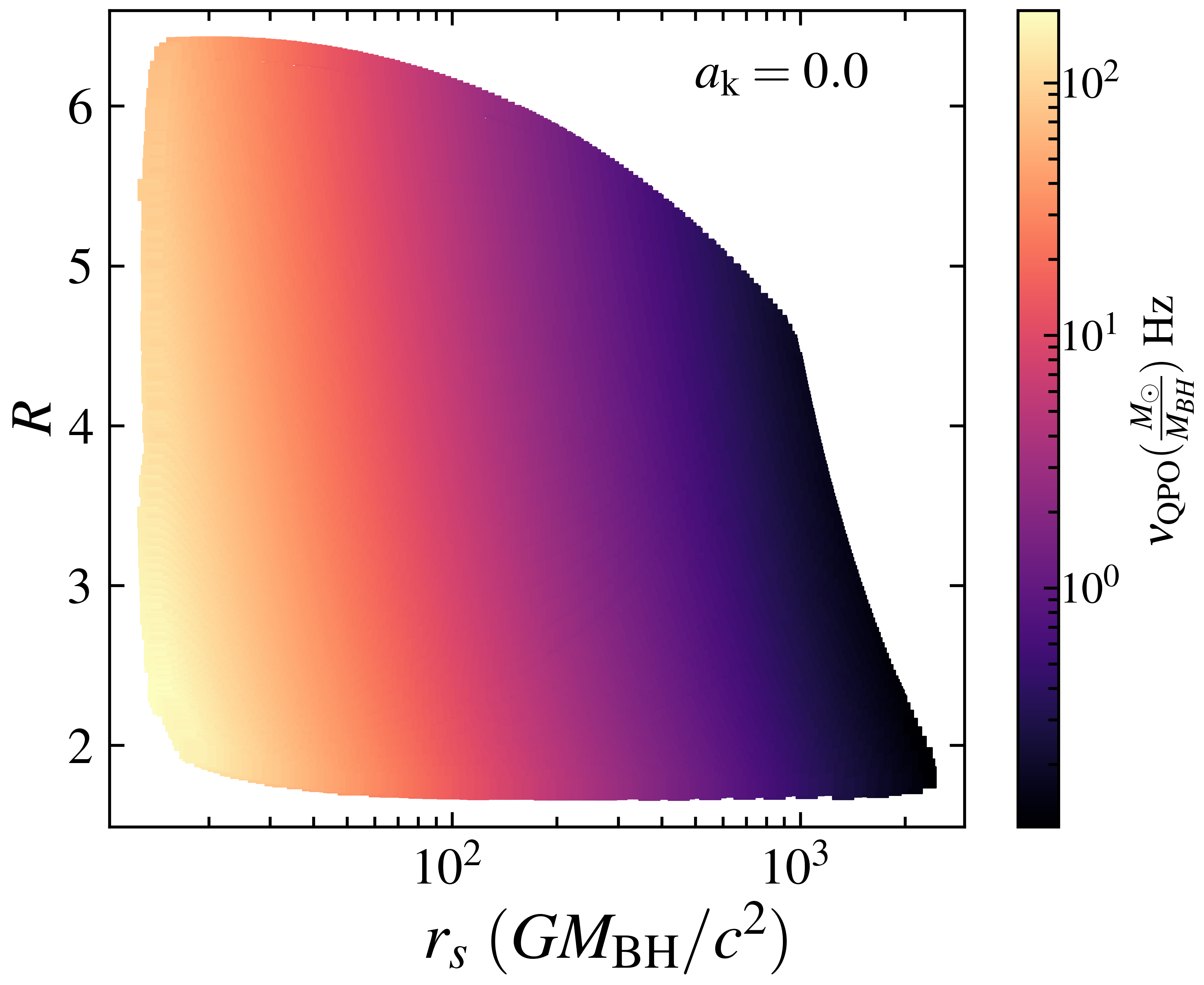}
       \vskip 0.5 cm
	   \includegraphics[width=\columnwidth]{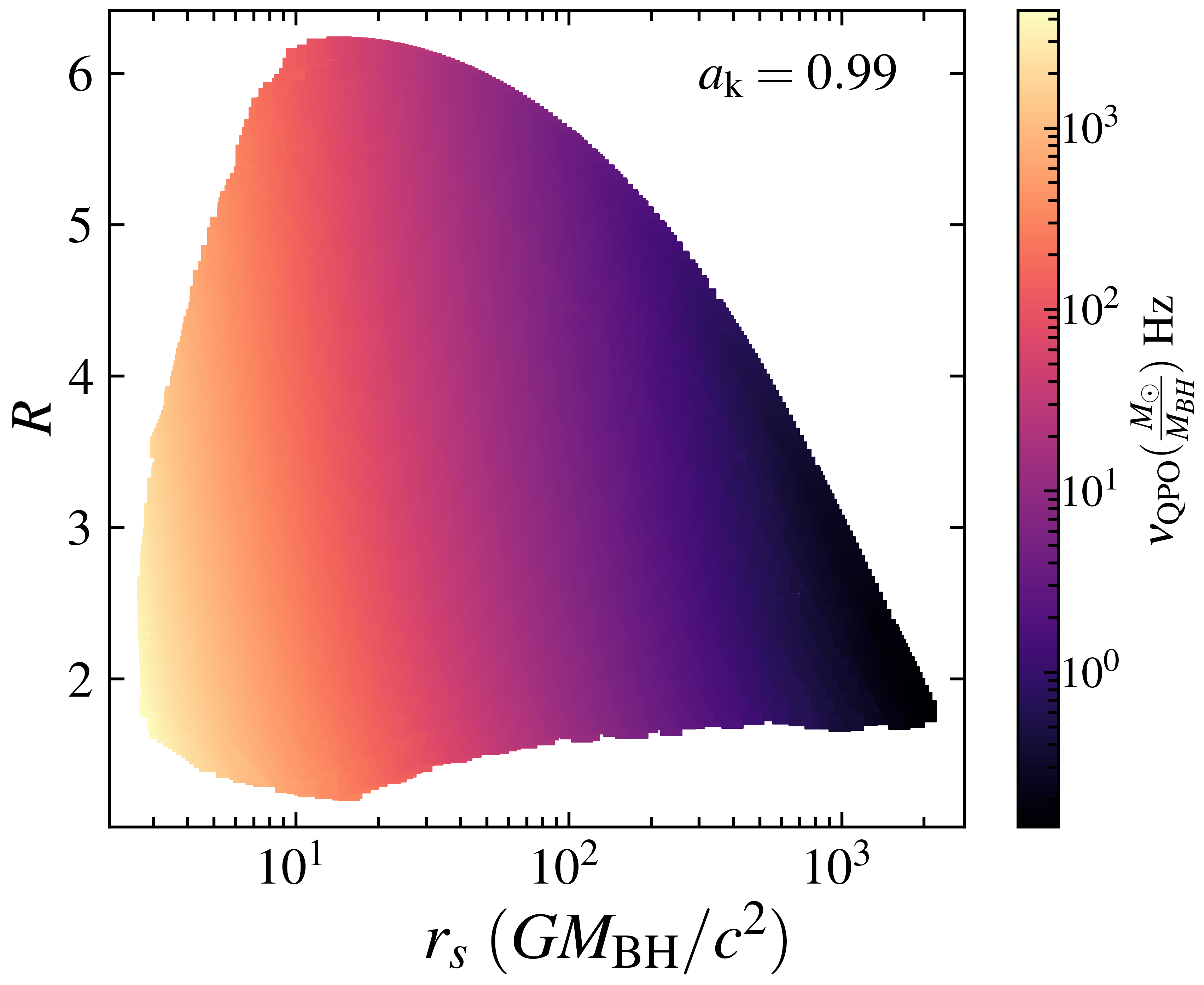}
	\end{center}
	\caption{Two-dimensional projection of the three-dimensional parameter space defined by the shock location ($r_{s}$), compression ratio ($R$), and QPO frequency $\nu_{\rm QPO}(M_\odot/M_{\rm BH})$, where the vertical color bar indicates the range of QPO frequencies on a logarithmic scale. Results are shown for weakly rotating ($a_{\rm k} = 0.0$, upper panel) and rapidly rotating ($a_{\rm k} = 0.99$, lower panel) black holes. See the text for details. 
    } 
    \label{fig:05}
\end{figure}

Within this framework, we consider that fluid elements in the post-shock region subsequently undergo nearly free-fall motion toward the black hole event horizon. The corresponding infall timescale of matter from the shock location to the horizon provides a natural dynamical timescale for the system. We therefore argue that the inverse of this infall timescale offers a robust estimate of the QPO frequency, establishing a direct correspondence between the observed quasi-periodic variability and the properties of the standing shock, along with the underlying global accretion flow parameters. Accordingly, following \citet{Chakrabarti-Manickam2000, Iyer-etal2015, Dihingia-etal2019}, we estimate the QPO frequency as,
\begin{equation}\label{nuqpo}
    \nu_{\rm QPO} = \frac{c/r_g}{(2)^{3/2}\pi R r_s \sqrt{r_s-2}}~{\rm Hz}.
\end{equation}
We compute $\nu_{\rm QPO}$ for all possible shock-induced global accretion solutions by varying $\mathcal{E}$ and $\lambda$ within the hybrid model for flows accreting onto weakly ($a_{\rm k}=0$, upper panel) and rapidly ($a_{\rm k}=0.99$, lower panel) rotating black holes. The results are presented in Fig. \ref{fig:05}, which shows a two-dimensional projection of the three-dimensional parameter space involving $r_{s}$, $R$, and $\nu_{\rm QPO}$. Since $\nu_{\rm QPO}$ depends directly on $r_{\rm s}$ and $R$, we recast the shock parameter space on the $r_{s}$–$R$ plane instead of the $\lambda$–$\mathcal{E}$ plane. Our formalism yields both strong ($R > 4$) and weak ($R \sim 1$) shocks. We further find that high-frequency QPOs generally occur when both $r_s$ and $R$ are small, whereas low-frequency QPOs arise when shocks form at larger radii with a wide range of $R$. These findings provide valuable insights for constraining model parameters to explain observations, as discussed next.

We note that, although distinct accretion solutions may produce similar shock locations, the corresponding QPO frequencies ($\nu_{\rm QPO}$) are generally expected to remain distinguishable owing to their combined dependence on the shock location ($r_{\rm s}$) and the shock compression ratio ($R$).

\begin{figure}
    \begin{center}
        \includegraphics[width=\columnwidth]{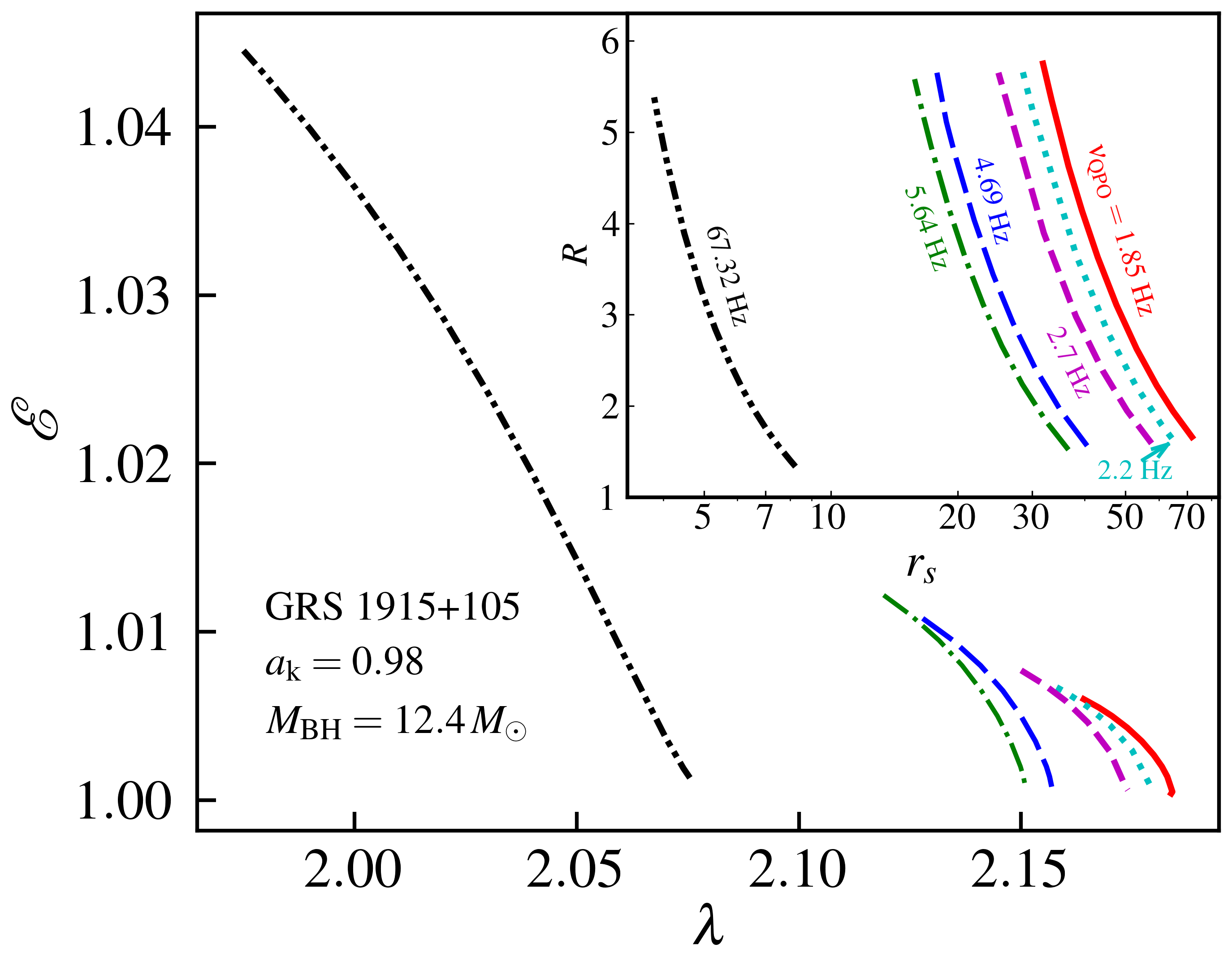}
    \end{center}   
    \caption{Allowed ranges of angular momentum ($\lambda$) and energy ($\mathcal{E}$) that satisfactorily reproduce both low-frequency and high-frequency QPOs ($\nu_{\rm QPO}$) observed in GRS~1915$+$105. The inset shows the corresponding allowed ranges of the shock location ($r_{s}$) and compression ratio ($R$). The solid (red), dotted (cyan), dashed (purple), long-dashed (blue), dot–dashed (green), and dot–dot–dashed (black) curves represent results corresponding to $\nu_{\rm QPO}=1.85$ Hz, $2.2$ Hz, $2.7$ Hz, $4.69$ Hz, $5.64$ Hz, and $67.32$ Hz, respectively. See the text for details.
    }
    \label{fig:06} 
\end{figure}

\begin{figure*}
    \begin{center}
        \includegraphics[width=\textwidth]{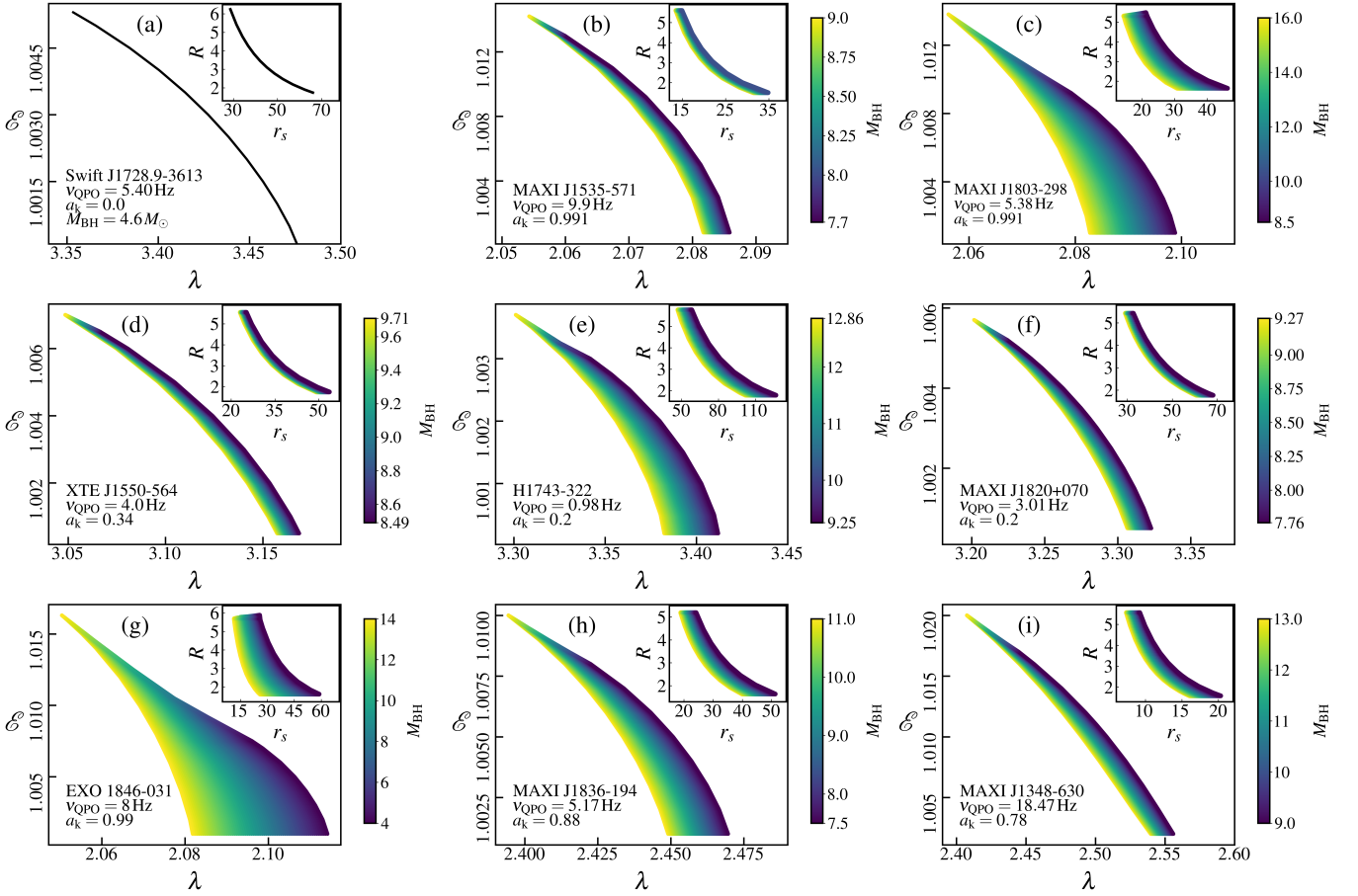}
    \end{center}
    \caption{Parameter space in $\lambda$–$\mathcal{E}$ plane and the corresponding $r_{s}$–$R$ plane for the nine BH–XRBs considered in this study (shown in nine panels) that reproduce the observed QPO centroid frequencies $\nu_{\rm QPO}$. Each panel is annotated with the source name, black hole mass, dimensionless spin parameter, and the observed QPO centroid frequency. Owing to uncertainties in the estimated black hole mass, $M_{\rm BH}$, the inferred solutions appear as finite-width bands rather than discrete curves. Color scale indicates the range of $M_{\rm BH}$. See the text for details. 
    }
    \label{fig:07}
\end{figure*}

\begin{table*}
	\caption{Physical and observable parameters of nine BH-XRBs under consideration. Columns 1$-$4 list the black hole mass ($M_{\rm BH}$), spin parameter ($a_{\rm k}$), and the observed QPO centroid frequency ($\nu_{\rm QPO}$). Columns~5 and~6 present the ranges of the shock location ($r_{s}$) and compression ratio ($R$) that successfully reproduce the observed QPO frequencies within our model, while the relevant references are provided in column~7. See the text for details.}
	\centering{
	\begin{tabular}{l c c c c c c c c c c l}
		\hline \hline
		Source \hspace{0.5 cm}
		& $M_{\rm BH}$ \hspace{0.5 cm}
		& $a_{\rm k}$ \hspace{0.5 cm}
        & $\nu_{\rm QPO}$ \hspace{0.5 cm}
		& $r_s$ \hspace{0.5 cm}
		& $R$ \hspace{0.5 cm}
        & References$^\ddagger$ \\
		
		Name \hspace{0.5 cm}
		& ($M_\odot$) \hspace{0.5 cm}
		&  \hspace{0.5 cm}
		&  (Hz) \hspace{0.5 cm}
		& ($GM_{\rm BH}/c^2$) \hspace{0.5 cm}
		& \hspace{0.5 cm}
        & \hspace{0.5 cm} \\
		\hline \hline

        Swift J1728.9$-$3613 \hspace{0.5 cm}
		& $4.6$ \hspace{0.5 cm}
		& $0.00$ \hspace{0.5 cm}
		& $5.40$ \hspace{0.5 cm}
		& \thickspace $28.59-66.01$ \hspace{0.5 cm}
		& \thickspace $1.74-6.24$ \hspace{0.5 cm}
        & 1, 1, 1\\

		MAXI J1535$-$571 \hspace{0.5 cm}
		& $7.7-9.9$ \hspace{0.5 cm}
		& $0.99$ \hspace{0.5 cm}
		& $9.90$ \hspace{0.5 cm}
		& \thickspace $13.46-34.86$ \hspace{0.5 cm}
		& \thickspace $1.50-5.63$ \hspace{0.5 cm}
        & 2, 3, 4\\

		MAXI J1803$-$298 \hspace{0.5 cm}
		& $8.5-16.0$ \hspace{0.5 cm}
		& $0.99$ \hspace{0.5 cm}
		& $5.38$ \hspace{0.5 cm}
		& \thickspace $14.05-46.26$ \hspace{0.5 cm}
		& \thickspace $1.51-5.48$ \hspace{0.5 cm}
        & 5, 6, 5\\

		XTE J1550$-$564 \hspace{0.5 cm}
		& $9.1 \pm 0.61$  \hspace{0.5 cm}
		& $0.34$  \hspace{0.5 cm}
		& $4.00$ \hspace{0.5 cm}
		& \thickspace $23.02-53.59$ \hspace{0.5 cm}
		& \thickspace $1.73-5.58$ \hspace{0.5 cm}
        & 7, 8, 9\\

		H 1743$-$322 \hspace{0.5 cm}
		& $11.21^{+1.65}_{-1.96}$ \hspace{0.5 cm}
		& $0.20$ \hspace{0.5 cm}
		& $0.98$ \hspace{0.5 cm}
		& \thickspace $46.43-126.62$ \hspace{0.5 cm}
		& \thickspace $1.73-5.86$ \hspace{0.5 cm}
        & 10, 11, 12\\

		MAXI J1820$+$070 \hspace{0.5 cm}
		& $8.48^{+0.79}_{-0.72}$\hspace{0.5 cm}
		& $0.20$  \hspace{0.5 cm}
		& $3.01$ \hspace{0.5 cm}
		& \thickspace $28.85-68.38$ \hspace{0.5 cm}
		& \thickspace $1.76-5.48$ \hspace{0.5 cm}
        & 13, 14, 15\\

		EXO 1846$-$031 \hspace{0.5 cm}
		& $9.0 \pm 5.0$  \hspace{0.5 cm}
		& $0.99$  \hspace{0.5 cm}
		& $8.00$ \hspace{0.5 cm}
		& \thickspace $11.36-58.76$ \hspace{0.5 cm}
		& \thickspace $1.47-5.87$ \hspace{0.5 cm}
        & 16, 17, 17\\

		MAXI J1836$-$194 \hspace{0.5 cm}
		& $7.5-11.0$  \hspace{0.5 cm}
		& $0.88$  \hspace{0.5 cm}
		& $5.17$ \hspace{0.5 cm}
		& \thickspace $18.58-51.23$ \hspace{0.5 cm}
		& \thickspace $1.62-5.31$ \hspace{0.5 cm}
        & 18, 19, 18\\
        
		MAXI J1348$-$630 \hspace{0.5 cm}
		& $11.0 \pm 2.0$  \hspace{0.5 cm}
		& $0.78$  \hspace{0.5 cm}
		& $18.47$ \hspace{0.5 cm}
		& \thickspace $7.38-20.34$ \hspace{0.5 cm}
		& \thickspace $1.58-5.57$ \hspace{0.5 cm}
        & 20, 21, 22\\
		
		\hline
	\end{tabular}
    }
    \begin{list}{}{}
    \item[$^\ddagger$]References for black hole mass ($M_{\rm BH}$), spin ($a_{\rm k}$) and  QPO frequency ($\nu_{\rm QPO}$) are given in column 7 in sequential order.
    \end{list}
    \justify
	\noindent{\bf References:} 
    1: \cite{Saha-etal2022}, 2: \cite{Chen-etal2022}, 3: \cite{Liu-etal2022}, 4: \cite{Shang-etal2019}, 5: \cite{Chand-etal2022}, 6: \cite{Feng-etal2022}, 7: \cite{Su-etal2015}, 8: \cite{Motta-etal2015}, 9: \cite{Orosz-etal2011}, 10: \cite{Chand-etal2020}, 11: \cite{Tursunov-etal2018}, 12: \cite{Molla-etal2017}, 13: \cite{Ma-etal2023}, 14: \cite{Guan-etal2021}, 15: \cite{Torres-etal2020}, 16: \cite{Liu-etal2021}, 17: \cite{Draghis-etal2020}, 18: \cite{Jana-etal2016}, 19: \cite{Reis-etal2012}, 20: \cite{Zhang-etal2020}, 21: \cite{Jia-etal2022}, 22: \cite{Lamer-etal2021}.\\
     
	\label{Tab:01}
\end{table*}
   
\section{ASTROPHYSICAL IMPLICATION}\label{sec:05}

In this section, we present theoretical estimates of key physical parameters, namely, energy ($\mathcal{E}$), angular momentum ($\lambda$), shock location ($r_{s}$), and compression ratio ($R$), which are inferred from the observed QPO frequency ($\nu_{\rm QPO}$) across several black hole X-ray binaries. Our primary focus is on low-frequency QPOs (LFQPOs).

We begin by focusing on the well known BH–XRB source GRS 1915$+$105, which exhibits a diverse range of QPOs associated with its various temporal variability classes. The presence of QPOs across several variability classes in this source has been reported by \citet{Athulya-etal2022}, providing valuable observational constraints for modelling the underlying accretion flow properties. Previous studies established the fundamental physical parameters of this source, including black hole mass of $12.4\,{\rm M}_\odot$ \citep{Reid-etal2014} and spin parameter of $a_{\rm k}=0.98$ \citep{Blum-etal2009}. Adopting these constraints, we utilise the observed QPO frequencies corresponding to different variability classes to delineate the allowed shock parameter space and to estimate the shock location ($r_{s}$), and compression ratio ($R$), as shown in Fig. \ref{fig:06}. In particular, we consider LFQPOs detected in the $\chi$, $\beta$, $\alpha$, $\rho$, and $\rho^{\prime}$ classes \citep{Valerie-etal2006, Athulya-etal2022}, along with the high-frequency QPO (HFQPO) observed in the $\gamma$ class \citep{Majumder-etal2025}, and aim to account for the corresponding QPO frequencies within our framework. In Fig. \ref{fig:06}, the solid (red), dotted (cyan), dashed (purple), long-dashed (blue), dot–dashed (green), and dot–dot–dashed (black) curves represent results corresponding to $\nu_{\rm QPO}=1.85$ Hz, $2.2$ Hz, $2.7$ Hz, $4.69$ Hz, $5.64$ Hz, and $67.32$ Hz, respectively. We find that the present model satisfactorily reproduces the observed $\nu_{\rm QPO}$ across different temporal variability classes over a broad range of the flow parameters $\lambda$ and $\mathcal{E}$. More specifically, LFQPOs are preferentially produced in flows with relatively higher angular momentum ($\lambda$), whereas HFQPOs arise at comparatively lower $\lambda$. In contrast, LFQPOs are confined to a narrower range of specific energy, while HFQPOs are obtained over a wider span of $\mathcal{E}$. Within the present formalism, QPOs originate from the aperiodic modulation of the shock front, with shock transitions occurring at larger radii for flows possessing higher angular momentum. Consequently, LFQPOs are associated with shocks forming farther away from the black hole. On the other hand, HFQPOs are yielded when the shock front undergoes oscillations at smaller radii closer to the event horizon, where relatively lower angular momentum flows combined with a broad range of energies naturally give rise to high frequency oscillations.

We extend this comprehensive analysis to an additional nine Galactic BH–XRBs, namely Swift J1728.9$-$3613, MAXI J1535$-$571, MAXI J1803$-$298, XTE J1550$-$564, H 1743$-$322, MAXI J1820$+$070, EXO~1846$-$031, MAXI J1836$-$194, and MAXI J1348$-$630. For all these sources, the black hole mass and spin are reasonably well constrained, and the corresponding observed QPO centroid frequencies ($\nu_{\rm QPO}$) are readily available in the literature. Utilizing these observational inputs, we identify the allowed ranges of the flow parameters $\mathcal{E}$ and $\lambda$, and subsequently determine the corresponding shock location ($r_{s}$) and compression ratio ($R$) that yield model predicted QPO frequencies consistent with the observed $\nu_{\rm QPO}$. The details of the selected sources are summarized in Table \ref{Tab:01}, where columns 1$-$4 list the source name, black hole mass ($M_{\rm BH}$), spin ($a_{\rm k}$), and observed QPO frequency ($\nu_{\rm QPO}$), respectively. Columns 5 and 6 present the ranges of $r_{s}$ and $R$ that theoretically reproduce the observed QPO frequencies, while the relevant references are provided in column 7.

In Fig.~\ref{fig:07}, we present the results obtained for these nine BH–XRBs in the $\lambda$–$\mathcal{E}$ parameter space. Owing to uncertainties in the black hole mass, the inferred parameter spaces appear as finite-width bands rather than as discrete curves. The inset in each panel further illustrates the corresponding allowed ranges of the shock location ($r_{s}$) and compression ratio ($R$) associated with the admissible $\lambda$–$\mathcal{E}$ parameter space. The color code denotes the range of source mass ($M_{\rm BH}$). For clarity, each panel is annotated with the source name, $M_{\rm BH}$, $a_{\rm k}$, and $\nu_{\rm QPO}$. Consistent with our results for GRS 1915$+$105, we find that the observed QPO frequencies for all sources can be satisfactorily reproduced within the hybrid disc geometry framework over a broad range of model parameters, including $\mathcal{E}$, $\lambda$, $r_{s}$, and $R$. These results, shown in Fig.~\ref{fig:07}, demonstrate that standing shock solutions within global accretion models provide a robust and unified framework for interpreting observational QPO constraints across a diverse sample of Galactic BH–XRBs.

\section{Conclusion}\label{sec:06}

In this study, we examine the extent to which semi-analytical global accretion solutions can reproduce the flow properties obtained from numerical simulations in the presence of shocks. A direct comparison reveals that no single semi-analytical disc geometry is able to adequately match the simulation profiles across the entire radial domain (see Fig. \ref{fig:02}). Instead, the pre-shock region of the flow exhibits closer agreement with the conical disc solution (Model-C), whereas the post-shock region is better described by the vertical equilibrium disc solution (Model-V), where enhanced thermal pressure drives significant vertical expansion while maintaining hydrostatic balance in the vertical direction.

Motivated by these complementary behaviors, we introduce a hybrid disc geometry in which the pre-shock accretion flow follows Model-C and the post-shock flow is governed by Model-V. The resulting shock-induced global accretion solution exhibits a close agreement with the simulation results in both dynamical and thermodynamical properties. In particular, the radial profiles of the Mach number ($M$) and temperature ($T$) predicted by the hybrid disc model closely match those from the simulation, and the inferred shock location differs by only $\sim 10\%$ from that obtained numerically (see Fig. \ref{fig:03}). The associated compression ratio across the shock is also found to be physically reasonable, further supporting the consistency of the model.

Overall, these results demonstrate that a hybrid disc geometry provides a physically motivated description of shocked accretion flows that is in agreement with numerical simulations. Within this framework, we obtain the shock parameters space in $\lambda-\mathcal{E}$ plane using the hybrid disc geometry for both weakly rotating ($a_{\rm k}\rightarrow 0$) and rapidly rotating ($a_{\rm k}=0.99$) black holes (see Fig. \ref{fig:04}). The proposed approach captures the essential features of both the pre- and post-shock regions and offers a reliable foundation for modelling shock driven temporal variability signatures in black hole accretion systems.

Within this framework, we explore the origin of QPOs in BH-XRBs in the context of global accretion flows harboring shock solutions. Using the hybrid disc geometry, we constrain the flow parameters, namely energy ($\mathcal{E}$) and angular momentum ($\lambda$) (as well as the derived quantities $r_{s}$ and $R$), that satisfactorily reproduce the observed QPO centroid frequencies ($\nu_{\rm QPO}$). Our analysis phenomenologically demonstrates that oscillations of the shock front provide a self-consistent mechanism for generating the observed QPOs in BH-XRBs (see Fig. \ref{fig:05}).

Considering GRS 1915$+$105 as a benchmark source with well constrained mass and spin, we show that the hybrid disc model successfully reproduces both LFQPOs and HFQPOs observed across different variability classes. We find that LFQPOs are preferentially associated with flows possessing relatively higher angular momentum with shocks forming at larger radii, whereas HFQPOs arise from lower angular momentum flows in which the shock forms closer to the black hole. Furthermore, LFQPOs are confined to a narrow range of flow energy, while HFQPOs span a broader energy domain. This finding highlights the distinct physical conditions governing these two classes of QPOs (see Fig. \ref{fig:06}).

Extending our analysis to nine additional Galactic BH-XRBs, we demonstrate that the present model formalism is capable to reproduce the observed QPO frequencies for these sources within the same theoretical framework. The inferred ranges of $\mathcal{E}$, $\lambda$, $r_{s}$, and $R$ are physically plausible and exhibit systematic trends that are broadly consistent across sources despite uncertainties in black hole mass estimates. These results establish that shocked solutions in global accretion flows provide a plausible and compelling interpretation of QPO phenomenology in BH-XRBs, reinforcing the possible role of shock dynamics in understanding accretion driven temporal variability (Fig. \ref{fig:07}).

Finally, we note the limitation of this study. The present model does not include high viscosity Keplerian disc component along with the sub‑Keplerian flow, as solving the coupled two‑component accretion equations introduces considerable mathematical complexity. In addition, the present framework does not account for alternative QPO mechanisms, such as misaligned disc precession \citep{Zhang-etal2020a} or GRMHD based variability \citep{Dihingia-Mizuno2025}. Nevertheless, the shock‑induced hybrid geometry adopted in this study provides a distinct dynamical mechanism in which QPOs arise from radial oscillations of the post‑shock region, yielding physical interpretations that differ fundamentally from those predicted by other QPO models.

\section*{Data Availability}

The data underlying this article will be available upon reasonable request.

\section*{Acknowledgment}

MS and SD acknowledge the support from the Department of Physics, IIT Guwahati, for providing the facilities to complete this work.

\bibliography{reference}

\appendix

\section{ Dependence of Shock Location on Black Hole Spin}\label{app:spin}

\begin{figure}
    \begin{center}
        \includegraphics[width=\columnwidth]{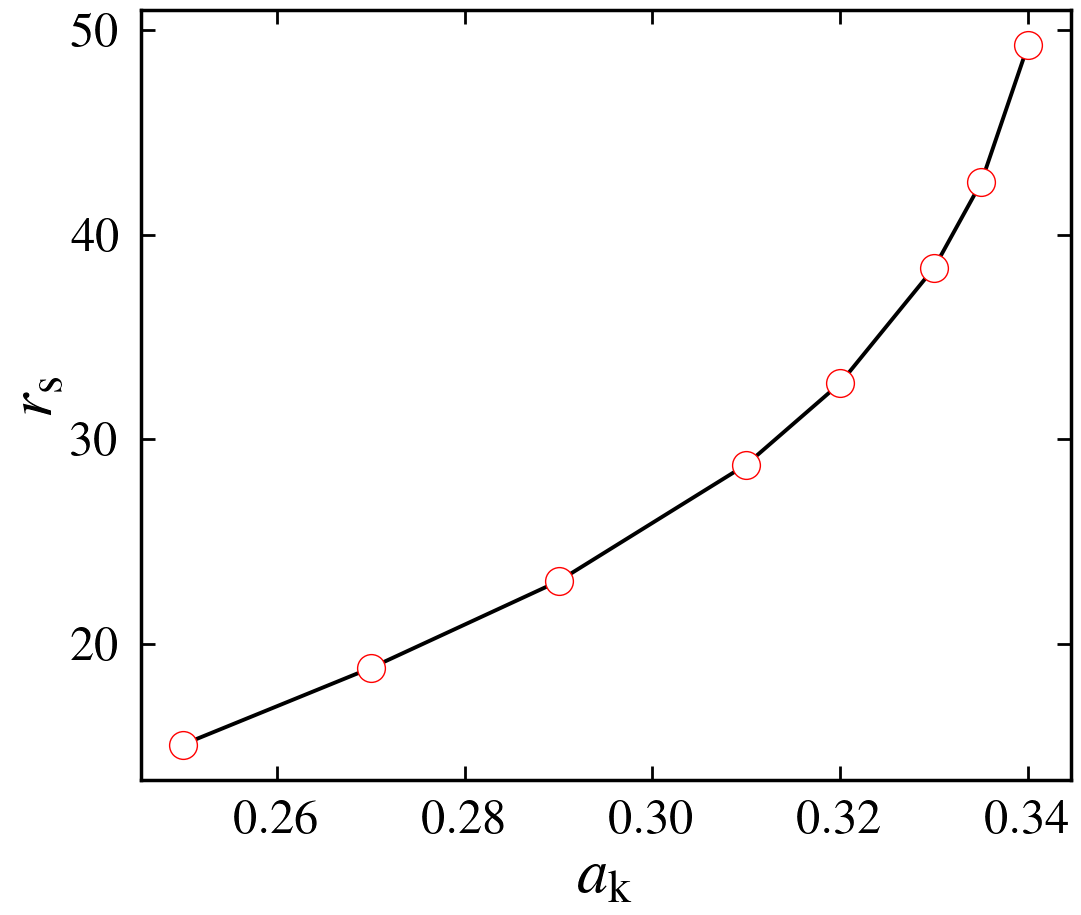}
    \end{center}
    \caption{ Variation of shock location ($r_{\rm s}$) with black hole spin($a_{\rm k}$). Here, we choose flow energy as $\mathcal{E} = 1.007$ and angular momentum as $\lambda=3.048$. See the text for details. 
    }
    \label{fig:app_spin}
\end{figure}

 The dependence of the shock location ($r_{\rm s}$) on the black hole spin ($a_{\rm k}$) is examined by keeping the flow energy and angular momentum fixed at $\mathcal{E}=1.007$ and $\lambda=3.048$, respectively, while varying the spin. The obtained results are presented in Fig.~\ref{fig:app_spin}.
%
We find that the shock location exhibits a systematic outward shift with increasing black hole spin which is consistent with the previous studies of shocked accretion flows around rotating black holes \cite[]{Mao-etal2025}. For the chosen flow parameters, the shock forms at a minimum radius of $\sim 15.06\,r_{\rm g}$ when $a_{\rm k}=0.25$. As the spin increases, the shock front gradually moves outward and reaches $r_{\rm s}\sim 49.28\,r_{\rm g}$ for $a_{\rm k}=0.34$, corresponding to a net displacement of $\Delta r_{\rm s}\sim 34.22\,r_{\rm g}$ over the considered spin range. This behaviour clearly demonstrates that, for a fixed set of flow parameters, increasing black hole spin shifts the shock front to larger radii.



\end{document}